# Entropic Diagram Characterization of Quantum Coherence: Degenerate Distillation and the Maximum Eigenvalue Uncertainty Bound

Tariq Aziz[1], Meng-Long Song[1], Liu Ye[1], Dong Wang[1,*]

[1] *School of Physics and Optoelectronic Engineering, Anhui University, Hefei 230601, China*

*Email: dwang@ahu.edu.cn

We present a geometric approach to characterize quantum coherence through entropic diagrams that map the eigenvalue spectra of density matrices into a two-dimensional space defined by von Neumann and Tsallis-2 entropies. We derive the relative entropy of coherence using the Schur–Horn majorization theorem and introduce a new coherence monotone—the relative cross-entropy of coherence—which is valid only for mixed states and whose behavior under various classes of quantum operations is examined in detail. Interestingly, our approach reveals distinct entropy-coherence distillation curves and introduces the notion of degenerate coherence distillation, where coherence is concentrated into higher-dimensional, partially uniform spectral configurations. Although the degenerate coherence distillation does not always surpass standard qubit coherence distillation in asymptotic rate, it may offer an operationally meaningful alternative for protocols that require high-dimensional or symmetrically structured coherence resources. In addition, we derive a refined entropic uncertainty relation that incorporates the maximum eigenvalue of post-measurement states, which leads to state-dependent bounds that improve upon the conventional Maassen–Uffink bound. These findings suggest that entropic diagrams may serve not only as a useful visual and analytical tool but also as a guide for developing resource-efficient strategies in quantum information processing.

## I. Introduction

Quantum coherence is a fundamental feature of quantum systems, which is naturally linked to the superposition principle and is responsible for many nonclassical phenomena in quantum optics and information theory. The concept was formulated through correlation functions to describe interference patterns in wave fields [1–3]. However, in modern quantum information science, quantum coherence is recognized as a valuable resource that can be quantified, manipulated, and consumed, similar to entanglement and other quantum resources [4–13]. The concept is relevant in many research fields such as quantum computation, metrology, and thermodynamics [14–19].

In general, a resource theory of quantum coherence has three foundational elements: the set of free (incoherent) states, a class of free operations that do not generate coherence from these states, and a set of valid coherence measures that quantify the resource while it behaves monotonically under free operations [20–21]. Over the past decade, several hierarchies of free operations have emerged, which are based upon different physical and mathematical considerations. Among these, maximally incoherent operations (MIO) encompass all completely positive trace-preserving maps that preserve incoherent states [4]. A more constrained subset, known as incoherent operations (IO), was introduced by Baumgratz *et al.* [5–6]; it requires each Kraus operator to map incoherent states to incoherent states. Building further, strictly incoherent operations (SIO) impose an even tighter constraint, which demands that both the Kraus operators and their adjoints preserve incoherence [6–7]. Further developments include dephasing-covariant incoherent operations (DIO), which commute with the dephasing channel, and physically incoherent operations (PIO), which incorporate incoherent ancillas and projective measurements [8–10]. In addition to the widely studied classes of aforementioned free operations, several alternative frameworks have been proposed to capture different physical aspects of coherence, including genuinely incoherent operations (GIO), fully incoherent operations (FIO), and translationally invariant operations (TIO) [10–12]. These classes form a nested and well-structured operational hierarchy [22]. For a more comprehensive comparison of these operational classes and their defining properties, see Refs. [20–22].

Quantification of coherence relies on measures which satisfy specific axioms such as faithfulness, monotonicity under free operations, strong monotonicity under selective measurements, and convexity [20–21]. Notable coherence measures include the relative entropy of coherence $C_r(\rho) = S(\Delta[\rho]) - S(\rho)$, where $\Delta[\rho]$ denotes the fully dephased version of $\rho$, and the $l_1$-norm of coherence $C_{l_1}(\rho) = \sum_{i,j}|\rho_{ij}| - 1$ [5]. These quantifiers depict different physical and mathematical aspects of coherence: $C_r(\rho)$ connects with coherence distillation and information-theoretic tasks, while $C_{l_1}(\rho)$ is simpler to compute and directly captures the magnitude of coherence in the off-diagonal entries of a density matrix.

Coherence manipulation under free operations has been rigorously analyzed using the mathematical machinery of majorization. For pure states, Du *et al.* [23] established that a transformation $|\psi\rangle \to |\varphi\rangle$ is possible under IO if and only if $\Delta[|\psi\rangle\langle\psi|] \succ \Delta[|\varphi\rangle\langle\varphi|]$. However, these results have incomplete proof, particularly for higher-dimensional systems. Chitambar and Gour [24–25] addressed these gaps by a refined interpretation of the IO and introduced SIO, which ensured that both Kraus operators and their duals preserve the set of incoherent states. This refinement emphasized the need for operational rigor in coherence theories. Zhu *et al.* [26] rigorously proved the majorization criterion for IO and SIO and demonstrated that the coherence monotones under these operations follow the same structural properties as entanglement monotones. They showed that coherence monotones, defined using real symmetric concave functions $f$ over the probability simplex $\mathcal{F}_{sc}$, follow the same structural properties as entanglement monotones. For pure states $|\psi\rangle$, the coherence monotone is given by $C_f(|\psi\rangle) = f\big(\mu(|\psi\rangle)\big)$, where $\mu(|\psi\rangle)$ represents the vector of eigenvalues in descending order, while for mixed states, coherence monotones generalize to $C_f(\rho) = \min_{\{p_j,\rho_j\}} \sum_j p_j C_f(\rho_j)$, where the minimization is taken over all pure state decompositions of $\rho = \sum_j p_j \rho_j$. This concave and symmetric construction reinforces the monotonicity and convexity of coherence measures under IO and SIO, therefore, it solidifies the theoretical framework of coherence as a resource.

Based on the majorization-based perspective, more work has developed analytical and numerical tools to analyze coherence quantification, transformations, and optimize distillation protocols [27–29]. More specifically, semidefinite programming (SDP) has emerged as a powerful method for the characterization of feasible operations and computation bounds for coherence conversion under different classes of operations, which include SIO, DIO, and PIO. The SDP techniques enable exact or approximate evaluations of one-shot distillation rates, robustness measures, and transformation probabilities, which offer practical routes to address scenarios where analytical criteria are either insufficient or intractable [29–30]. Several researchers have investigated the process of coherence distillation, in which the purpose is to extract pure coherent states from a supply of mixed states by using allowed free operations. Winter and Yang [6] introduced operational coherence distillation rates under IO and MIO, while Regula *et al.* [31] used the SDP hierarchies to optimize distillation in general settings. The probabilistic, deterministic, assisted, and catalytic forms of distillation have also been explored [32–37], which reveal the rich structure of coherence dynamics in finite and asymptotic regimes.

On the other hand, the relation between entropy and the index of coincidence and error probability have been studied in earlier works [38–39], where analytical bounds and geometric insights were developed with applications in cryptography and rate-distortion theory. In a different context, Aiello and Woerdman [40] investigated constraints on polarization entropy and depolarization indices arising from light scattering, and introduced universal boundaries in the entropy–depolarization plane. Later, Tariq *et al.* [41–42] proposed a geometric framework based on the purity and depolarization indices to describe polarized light scattering. Their approach recovers the entropy–depolarization representation through indices of purity, which offers a simplified method for analyzing scattering behavior that may prove useful in biomedical optics and radar polarimetry. More recent contributions [43–44] have used information diagrams to refine entropic uncertainty relations, particularly for quantum measurements, which involve mutually unbiased bases and symmetric informationally complete POVMs. These techniques can potentially improve the identification of nonclassical features such as entanglement and steerability. Taken together, these developments illustrate how entropic diagrams provide a unifying framework for studying uncertainty, randomness, and structure in classical and quantum settings. These approaches may offer improved methods to detect non-classical correlations such as entanglement and steerability. Together, these studies present the usefulness of entropic diagram to analyze uncertainty, randomness, and structural properties across classical and quantum information science and optics. Nonetheless, their use in systematically characterizing quantum coherence has yet to be fully explored.

In this work, we introduce a geometric framework to characterize quantum coherence. We employed the Schur–Horn majorization theorem, which connects the spectral and diagonal representations of Hermitian operators and implies that the vector of diagonal entries is always majorized by the eigenvalue vector. We then construct entropy diagrams that reveal coherent structure and resource behavior in a visually intuitive way by mapping the eigenvalue spectra of quantum states into a two-dimensional space spanned by von Neumann entropy $S(\rho)$ and Tsallis-2 entropy $S_2(\rho)$. Unlike conventional coherence measures that rely on a single entropy functional, the entropic diagram captures the joint behavior of spectral distribution and coherence potential across the full state space, which may offer a more nuanced and intuitive representation of coherence.

From this entropic embedding, we introduce the concept of degenerate coherence distillation. Rather than converting states to the standard maximally coherent qubit, we consider distillation into a high-dimensional, symmetrically structured target state $\rho_{\text{deg}}$, whose spectrum is of the form $\lambda_1 \geq \lambda_2 = \cdots = \lambda_d$. These states lie along the upper boundary curve in the $S_2(\rho) - S(\rho)$ diagram that contains maximal entropy for a given Tsallis-2 value under the Schur-concavity of von Neumann entropy. This is different from the standard coherence distillation curves, which typically concentrate weight in the largest two eigenvalues. Degenerate distillation offers an alternative strategy that may be more practical in high-dimensional tasks such as Fourier transform protocols, qudit-based quantum computing, or optical mode multiplexing. Although this approach may not always outperform qubit-based distillation in terms of asymptotic rate, it provides operationally meaningful pathways when structured high-dimensional coherence is preferred.

Additionally, our framework enables the derivation of a refined entropic uncertainty relation (EUR) based on the largest eigenvalue. We show that the inequality $S(\rho) \geq 1 - \lambda_{\max}(\rho)$, holds for arbitrary states and extends to measurement outcomes, which yields a state-dependent EUR that may sometimes tighten the classical Maassen–Uffink bound [45].

Taken together, our contributions provide a systematic and visualizable toolset for analyzing coherence as a resource. By integrating geometric insights, majorization structure, and entropy-based measures, we establish new coherence monotones, uncover distillation pathways beyond the qubit model, and refine EUR for quantum systems. These findings may have potential applications in quantum communication, quantum metrology, and photonic technologies where control over structured coherence plays a critical role.

## II. The Suhr-Horn Majorization and coherence monotones

The mathematical foundation of coherence measures like relative entropy of coherence is deeply connected to the SH majorization theorem which is associated with the symplectic geometry through the Atiyah-Gullemin-Sternberg (AGS) convexity theorem, which asserts that the image of a compact connected symplectic manifold under a moment map is a convex polytope [46–47]. This result has profound implications in quantum mechanics, as it connects the symplectic structure of Hermitian matrices to convex geometric objects, which may allow us to describe the eigenvalue distribution of quantum states geometrically. The moment map associates a Hermitian matrix $A \in \mathcal{H}(d)$ to its diagonal matrix under the action of the maximal torus $T \subset U(d)$ captures the relationship between its spectral and diagonal components. The coadjoint orbits in the unitary Lie group $U(d)$ play a central role in this framework. For a Hermitian matrix, or more specifically, a $d-$dimensional density matrix $\rho$ with eigenvalue vector $\boldsymbol{\lambda}$ where $\lambda_1 \geq \lambda_2 \geq \cdots \geq \lambda_d$, the coadjoint orbit $\mathcal{O}_{\boldsymbol{\lambda}}$ of $\rho$ under the action of $U(d)$ consists of all matrices unitarily equivalent to $\rho$. The projection of its orbit onto the diagonal matrices, via the moment map, forms a convex polytope. This projection is governed by majorization relations, which establishes a direct link to the SH theorem. Specifically, the convex polytope formed by the diagonals of matrices in $\mathcal{O}_{\boldsymbol{\rho}_{ii}}$ is characterized by a condition $\boldsymbol{\rho}_{ii} \preccurlyeq \boldsymbol{\lambda}$, where $\boldsymbol{\rho}_{ii}$ is the vector of diagonal elements of a density matrix $\rho_{11} \geq \rho_{22} \geq \cdots \geq \rho_{dd}$, which means that,

$$\sum_{i=1}^{k} \lambda_i^{\downarrow} \geq \sum_{i=1}^{k} \rho_{ii}^{\downarrow}, \qquad \forall\, k, \tag{1}$$

where symbol $\downarrow$ represents the descending order, with equality for $k = d$. The matrix of diagonal elements $\rho_{\text{diag}}$ with zero off-diagonal elements is majorized by $\rho$, which means that $\rho_{\text{diag}} = \sum_j p_j U_j \rho U_j{}^{\dagger}$, where $U_j$ are unitary operators $\{p_j\}$ are probabilities. By the concavity of the von Neumann entropy of the decohered state $S(\rho_{\text{diag}})$, i.e., $S\left(\sum_j p_j \rho_{\text{diag}_j}\right) \geq \sum_j p_j S(\rho_{\text{diag}_j})$ [48], we have $S(\rho_{\text{diag}}) \geq \sum_j p_j S(U_j \rho U_j{}^{\dagger}) = \sum_j p_j S(\rho) = S(\rho)$, as it is invariant under unitary transformation and the probabilities sum to 1 ($\sum_j p_j = 1$). Hence, $S(\rho_{\text{diag}}) \geq S(\rho) \Rightarrow C_r(\rho) = S(\rho_{\text{diag}}) - S(\rho) \geq 0$, which is the relative entropy of coherence [5]. The non-negativity of $C_r(\rho)$ can also be interpreted through the lens of majorization, because $\rho_{\text{diag}} \preccurlyeq \rho$ the entropy $S(\rho_{\text{diag}})$ is guaranteed to be larger than $S(\rho)$, which ensures $C_r(\rho) > 0$.

Furthermore, one may define an alternative yet closely related coherence measure, the relative cross-entropy of coherence $C_r{}^{\text{cross}}(\rho)$ that shows cross-entropic differences between the diagonalized and full forms of $\rho$. Explicitly, we define it as the difference between two cross-entropy-like terms as

$$C_r{}^{\text{cross}}(\rho) = \text{Tr}[\rho \log(\rho_{\text{diag}})] - \text{Tr}[\rho_{\text{diag}} \log(\rho)]. \tag{2}$$

Interestingly, this definition is equivalent to defining a coherence measure as the reverse relative entropy of coherence. Eq. (02) can be expanded and rearranged as

$$C_r{}^{\text{cross}}(\rho) = \text{Tr}\left[\rho_{\text{diag}} \log(\rho_{\text{diag}})\right] - \text{Tr}\left[\rho_{\text{diag}} \log(\rho)\right] + \text{Tr}\left[(\rho - \rho_{\text{diag}}) \log(\rho_{\text{diag}})\right]. \tag{3}$$

Since the state $\rho_{\text{diag}}$ is diagonal by construction, and the operator $(\rho - \rho_{\text{diag}})$ has strictly zero diagonal elements, it follows immediately that $\text{Tr}\left[(\rho - \rho_{\text{diag}}) \log(\rho_{\text{diag}})\right] = 0$. Thus, the relative cross-entropy of coherence simplifies to,

$$C_r{}^{\text{cross}}(\rho) = \text{Tr}\left[\rho_{\text{diag}}\left(\log\left(\rho_{\text{diag}}\right) - \log \rho\right)\right]. \tag{4}$$

Recognizing this final expression, we identify it as the quantum relative entropy of $\rho_{\text{diag}}$ with respect to $\rho$, i.e.,

$$C_r{}^{\text{cross}}(\rho) = S\left(\rho_{\text{diag}}||\rho\right), \tag{5}$$

which can now be viewed as a reversed counterpart to the standard relative entropy of coherence $C_r = S\left(\rho||\rho_{\text{diag}}\right)$. Moreover, from the Klein's inequality, $C_r{}^{\text{cross}}(\rho)$ satisfies faithfulness (condition C1 Ref. [21]), which means $C_r{}^{\text{cross}}(\rho) = S\left(\rho_{diag}||\rho\right) \geq 0$.

In addition, $C_r{}^{\text{cross}}(\rho)$ satisfies criterion C2 [21], when interpreted with respect to operations that commute with the dephasing map. Specifically, if the quantum operation $\Lambda$ satisfies $(\Lambda(\rho))_{\text{diag}} = \Lambda\left(\rho_{\text{diag}}\right)$, then the relative cross-entropy of coherence obeys

$$\begin{aligned} C_r{}^{\text{cross}}\left(\Lambda(\rho)\right) &= S\left(\left(\Lambda(\rho)\right)_{\text{diag}}||\Lambda(\rho)\right) \\ &\leq S\left(\Lambda\left(\rho_{\text{diag}}\right)||\Lambda(\rho)\right) \\ &\leq S\left(\rho_{\text{diag}}||\rho\right) \\ &= C_r{}^{\text{cross}}(\rho). \end{aligned} \tag{6}$$

The inequality follows from the data-processing inequality for quantum relative entropy. This commutation condition holds for DIO, and also for SIO which are a subset of both DIO and IO [8,21]. Therefore, criterion C2 is satisfied within these operational frameworks, though not necessarily under all incoherent operations.

For strong monotonicity condition C3 [21], one needs a decomposition of the post-measurement state in terms of probabilities $\{p_n\}$ from measuring $\rho$. The problem is that the condition for the $C_r{}^{\text{cross}}(\rho)$ follows from the inequality under quantum Kraus operators $K_n$,

$$S(\rho_{\mathrm{diag}}||\rho) \geq \sum_n q_n S\left(\frac{K_n\rho_{\mathrm{diag}}K_n{}^{\dagger}}{q_n} \| \frac{K_n\rho K_n{}^{\dagger}}{p_n}\right), \tag{7}$$

where $q_n = \mathrm{Tr}\left(K_n\rho_{\mathrm{diag}}K_n{}^{\dagger}\right)$ and $p_n = \mathrm{Tr}\left(K_n\rho K_n{}^{\dagger}\right)$, whereas strong monotonicity(C3) requires $C_r{}^{\mathrm{cross}}(\rho) \geq \sum_n p_n C_r{}^{\mathrm{cross}}\left(\frac{K_n\rho K_n{}^{\dagger}}{p_n}\right)$. However, Eq. (7) involves $q_n$, which generally differs from $p_n$. Since there is no guarantee that $\sum_n q_n S\left((\rho_n)_{\mathrm{diag}}\|\rho_n\right) \leq \sum_n p_n S\left((\rho_n)_{\mathrm{diag}}\|\rho_n\right)$, the $C_r{}^{\mathrm{cross}}(\rho)$ may not necessarily satisfy strong monotonicity under incoherent operations. Moreover, it violates uniqueness condition C5 [21]; for a pure state $\rho = |\psi\rangle\langle\psi|$, $\rho_{\mathrm{diag}}$ is mixed, which leads to $S(\rho_{\mathrm{diag}}||\rho) \longrightarrow \infty$, due to singularity of $\log\rho$ [49]. This conflicts with the requirement that pure-state coherence measures be finite. Nonetheless, $C_r{}^{\mathrm{cross}}(\rho)$ satisfies the remaining conditions C4 (convexity) and C6 (additivity) due to the joint convexity of the relative quantum entropy and the linearity of the diagonal map ($(\sum_i p_i\,\rho_i)_{\mathrm{diag}} = \sum_i p_i\,(\rho_i)_{\mathrm{diag}}$):

$$C_r{}^{\mathrm{cross}}(\textstyle\sum_i p_i\,\rho_i) \leq \sum_i p_i\, C_r{}^{\mathrm{cross}}(\rho_i), \tag{8}$$

and additivity under tensor products:

$$C_r{}^{\mathrm{cross}}(\rho\, \otimes\, \sigma) = C_r{}^{\mathrm{cross}}(\rho) + C_r{}^{\mathrm{cross}}(\sigma), \tag{9}$$

as expected for a quantum relative entropy.

We also note an important limitation of the $C_r{}^{\mathrm{cross}}(\rho)$. For any pure state $\rho = |\psi\rangle\langle\psi|$ such that $\rho \neq \rho_{\mathrm{diag}}$, then $\mathrm{rank}(\rho) = 1$, whereas $\mathrm{rank}\left(\rho_{\mathrm{diag}}\right) \geq 2$. In such cases, $\mathrm{supp}\left(\rho_{\mathrm{diag}}\right) \nsubseteq \mathrm{supp}(\rho)$, which implies that $C_r{}^{\mathrm{cross}}(\rho) = +\infty$, due to the divergence of $\mathrm{Tr}\left[\rho_{\mathrm{diag}}\log\rho\right]$. Consequently, while $C_r{}^{\mathrm{cross}}(\rho)$ is well-defined and operationally meaningful for mixed states, its divergence for pure states limits its scope and prevents it from serving as a universal coherence measure. A similar issue was identified in the context of relative entropy of entanglement [49]. Ideally, a coherence measure should yield finite values for all states, even if it does not fulfill every standard axiom, such as the uniqueness condition (C5) [21]. For instance, coherence quantifiers based on Tsallis entropy demonstrate that finite measures are achievable without meeting all criteria [50-51]. Therefore, $C_r{}^{\mathrm{cross}}(\rho)$ cannot be considered a general coherence measure on the full Hilbert space. However, it remains well-defined and operationally meaningful for mixed states, particularly in contexts where divergence is avoided (e.g., in resource theories restricted to full-rank states or under specific noise models). Nonetheless, the divergence issue of $C_r{}^{\mathrm{cross}}(\rho)$ can be effectively addressed through several regularization strategies. These might include ε-

smoothing methods that ensure finiteness [52], operational smoothing via subchannel discrimination [53], and asymptotic regularization techniques developed in dynamical resource theories [54], where coherence is quantified through limits over channel repetitions. These or similar approaches that can be adapted to regularize ${C_r}^{\text{cross}}(\rho)$ to avoid divergence for pure states. Future work could explore regularization techniques or hybrid measures to address this limitation.

We further generate random density matrices $\rho$ in a $d$-dimensional Hilbert space by first drawing a complex $d \times d$ matrix M from the Ginibre ensemble, where each entry is an independent complex Gaussian random variable. A normalized density matrix is then constructed via $\rho = \mathrm{MM}^{\dagger}/\mathrm{Tr}(\mathrm{MM}^{\dagger})$, which corresponds to a normalized Wishart matrix and provides a widely used method for sampling random quantum states according to the Hilbert–Schmidt measure. We subsequently apply a set of Kraus operators $\{K_k\}$, chosen to be diagonal in the reference basis, to implement an incoherent quantum operation. The state transforms according to $\rho \mapsto \sum_k K_k \rho {K_k}^{\dagger}$ with renormalization applied if necessary to maintain unit trace. Numerical graphs of the original versus transformed coherence values in Figs. 1 (a), (c), and (d) show the conditions C2 (monotonicity), C4 (convexity), and C6 (additivity) hold for ${C_r}^{\text{cross}}(\rho)$, while some small violations of C3 (strong monotonicity) can be seen in Fig. 1 (b). These violations may arise in cases where coherence contributions interfere in a way that leads to $p_n < q_n$, though this would require a special form of $K_n$ that suppresses diagonal elements while it amplifies coherence effects. Such occurrences are rare under typical incoherent operations. It may be noted that while our theoretical justification of monotonicity is limited to operations that commute with the dephasing map (i.e., DIO and SIO), our numerical simulations show that ${C_r}^{\text{cross}}(\rho)$ also respects the relevant C2 condition under general IO, with no violations of monotonicity observed in the sampled instances. These results illustrate how, in practice, most axioms are met while subtle breakdowns of strong monotonicity may occur under certain random Kraus constructions.

A useful way to view ${C_r}^{\text{cross}}(\rho)$ is that compares the classical version of $\rho$ to $\rho_{\text{diag}}$ itself, which captures how costly it is—under relative entropy—to build up the off-diagonal (coherent) structure from a diagonal baseline for mixed states. Physically, it therefore, quantifies how much of a resource the off-diagonal elements represent compared to the purely diagonal state, with an immediate operational meaning in tasks where one tracks classical versus quantum features through a common processing channel. While $C_r(\rho)$ measures how distinguishable $\rho$ is from its classicalized counterpart $\rho_{\text{diag}}$, ${C_r}^{\text{cross}}(\rho)$ instead quantifies how much information is lost when $\rho_{\text{diag}}$ is used as an approximation to $\rho$, which may make it a more direct measure of coherence degradation, though it remains valid only for mixed states.

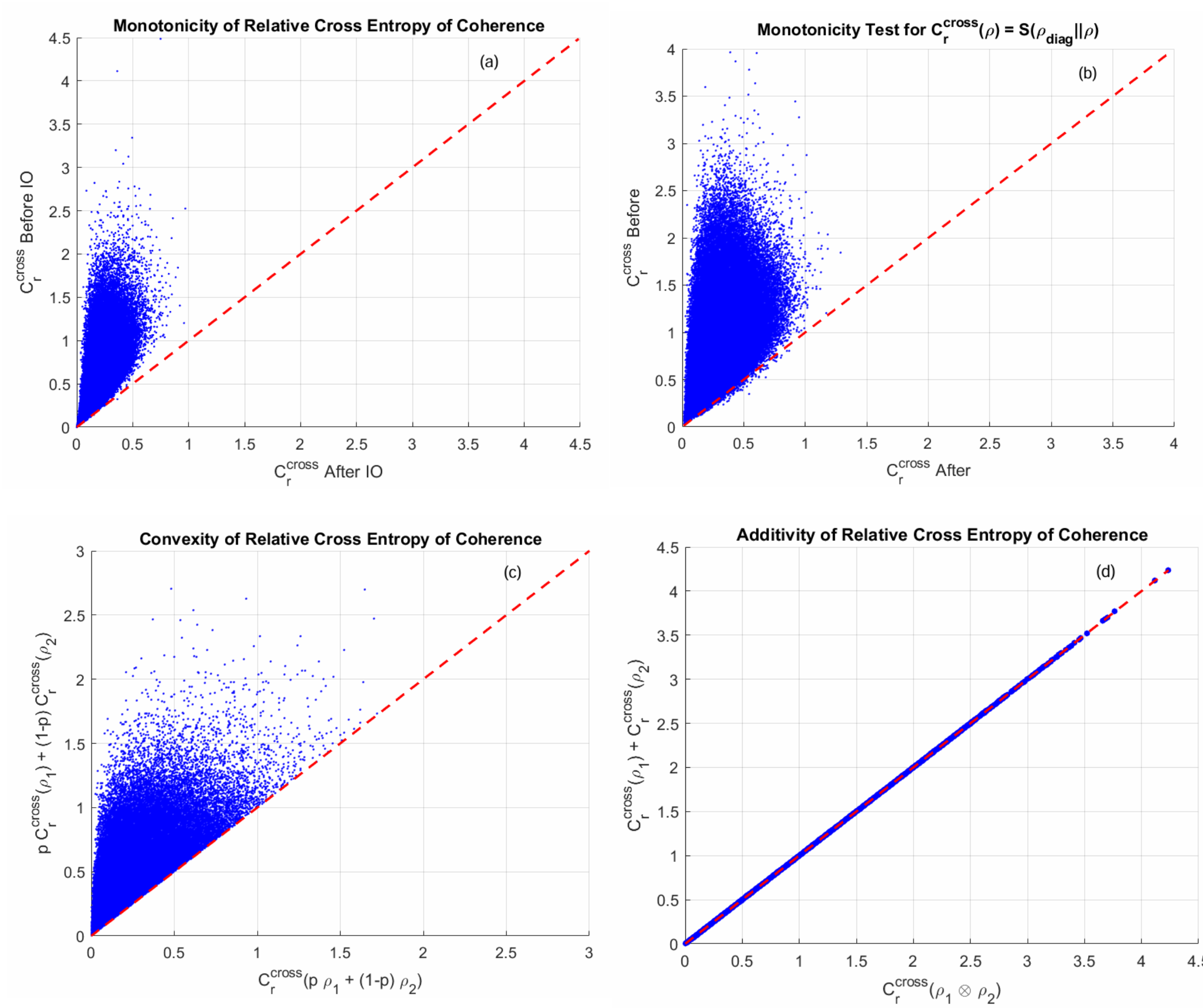

*FIG. 1 (a-d) Numerical analysis of the relative cross-entropy of coherence $C_r{}^{\text{cross}}(\rho)$ for an ensemble of randomly generated quantum states. (a) Monotonicity under incoherent operations: coherence after IO remains below or equal to the initial value. (b) Strong monotonicity test: most points satisfy the condition, with rare small violations indicating the subtle breakdown of strong monotonicity. (c) Convexity check: $C_r{}^{\text{cross}}(\rho)$ of a probabilistic mixture lies below the weighted sum of individual coherences, verifying convexity. (d) Additivity under tensor product: total coherence equals the sum of component coherences, with all data points aligning tightly with the identity line*.

Although the squared version of the SH majorization is not explicitly discussed in the general literature, we have observed that for density matrices, the squaring operation does not violate the majorization condition. We used a Monte Carlo framework to test the squared SH majorization condition on random density matrices generated using Haar-random unitary matrices, Schmidt decomposition, and eigenvalue-based constructions. Small perturbations were applied to ensure robustness while maintaining Hermiticity, positivity, and trace-1 properties. Across all tests, involving comparisons of squared eigenvalues and diagonal elements, no violations were observed, which confirms the numerical validity of the squared majorization condition. While the conjecture on the squared version of the SH majorization remains to be formally established, our findings suggest its validity in the context of density matrices. As a consequence, this opens up the possibility of deriving coherence monotones. Specifically, using the squared SH majorization, we can obtain both the partial and full relative Renyi and Tsallis of coherence for $q = 2$, which are defined as distinguishability measures $D_q(\rho||\rho_{\text{diag}})$, that may not exhibit strong monotonicity under incoherent operations [21, 50]. The squared eigenvalue structure is tied closely to purity such as in the trade-off coherence-purity relation [50]. Consequently, universal bounds between the von Neumann entropy and the Tsallis-2 (since Renyi-2 $\geq$Tsallis-2, we chose Tsallis-2) functionals can be derived, which may reflect the interplay of spectral spread and coherence in a quantum state. These bounds provide a systematic way to characterize coherence. Therefore, it may offer a more geometric understanding of coherence, which we describe in the next section.

## III. Entropic diagram characterizing coherence and the distillation curves

We begin by considering the set of valid eigenvalue vectors $\{\lambda_i\}_{i=1}^{d}$ of $d$ −dimensional density matrix $\rho$. Mathematically, these vectors lie in the standard probability simplex $\Delta^{d-1} \subset \mathbb{R}^d$, defined by $\lambda_i \geq 0$ and $\sum_{i=1}^{d} \lambda_i = 1$. Although $\Delta^{d-1}$ is itself a convex polytope with flat boundaries, we do not retain that characterization once we map each $\{\lambda_i\}$ to the two functionals von Neuman entropy $S(\rho)$ and Tsallis-2 entropy $S_2(\rho)$, which belong to the one side of the SH majorizations when $k = d$, because the transformation $\lambda \mapsto (S_2(\rho), S(\rho))$ is not linear. Consequently, the image of $\Delta^{d-1}$ in $\mathbb{R}^2$ becomes a curved region rather than a simplex. Moreover, by imposing additional constraints on the eigenvalue distribution (see Table. 1), we restrict ourselves to submanifolds of $\Delta^{d-1}$, which emerge as points, boundary arcs or distinct clusters in the $S_2(\rho) - S(\rho)$ plane. Physically, $S_2(\rho)$ rises as the eigenvalues become more uniform, which vanishes for a pure state and reaches $1 - 1/d$ for the maximally mixed state, while $S(\rho)$ increases from 0 to $\log(d)$. Collecting all sampled points thus reveals how specific eigenvalue pattern carve out recognizable quantum state as shown in Fig. 3.

In addition to these coherence-entropy mappings, we compare $C_r(\rho)$ and the $C_r{}^{\text{cross}}(\rho)$. Fig. 2 (a), we present a scatter plot of $C_r{}^{\text{cross}}(\rho)$ versus $C_r(\rho)$ for an ensemble of randomly generated four-dimensional quantum states. A color gradient encodes the $l_1$ −norm of coherence $C_{l1}(\rho)$, which reveals correlations with coherence content. Interestingly, while the qubit case is known to obey $C_r{}^{\text{cross}}(\rho) \geq C_r(\rho)$ for all states, our results for the qudit with $d = 4$ show that some of the points lie slightly above the identity line $y = x$, which indicates instances where $C_r{}^{\text{cross}}(\rho) \leq C_r(\rho)$. These small but systematic violations indicate that the ordering between coherence measures is dimension-dependent and not preserved in general. Such observations point toward subtle structural differences in coherence quantification and may inspire future work on resource-theoretic inequalities or coherence-based uncertainty relations in higher-dimensional quantum systems.

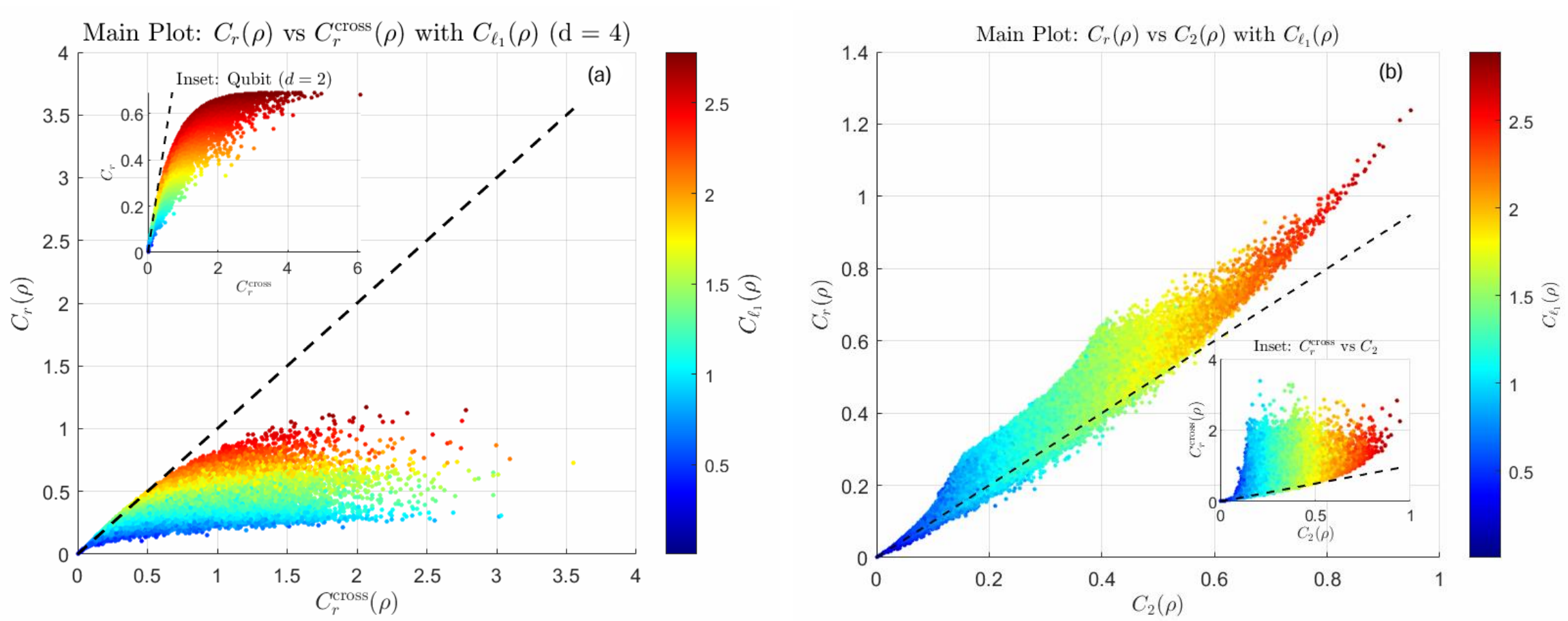

*FIG.2: (a) Comparison of $C_r(\rho)$ and $C_r{}^{\text{cross}}(\rho)$ for $d = 4$, with $C_{l1}(\rho)$ shown as a color gradient; inset: qubit case $d = 2$. (b) Comparison of $C_r(\rho)$ and $C_2(\rho)$ for $d = 4$, with inset showing $C_r{}^{\text{cross}}(\rho)$ vs. $C_2(\rho)$; color encodes $C_{l1}(\rho)$.*

One could alternatively employ the measure $C_2(\rho) = \sum_j \sqrt{\langle j|\rho^2|j\rangle} - 1$ in place of the Tsallis-2 entropy, as it satisfies all the fundamental axioms required in the resource theory of quantum coherence—such as strong monotonicity and convexity—unlike the original Tsallis relative entropy [51]. To explore its structural behavior, In Fig. 2(b), we visualize this measure by plotting $C_r(\rho)$ along the vertical axis against $C_2(\rho)$ on the horizontal axis, with the color gradient encoding the $C_{l1}(\rho)$. The inset displays a similar scatter plot of $C_r{}^{\text{cross}}(\rho)$ versus $C_2(\rho)$, which provides a comparative perspective. Both plots show that most quantum states satisfy $C_r(\rho) \geq C_2(\rho)$ and $C_r{}^{\text{cross}}(\rho) \geq C_2(\rho)$, which suggest that $C_2(\rho)$ tends to provide a lower bound on coherence relative to entropic measures. Nevertheless, these inequalities are not universally satisfied, with violations that appear in low-coherence regions. Moreover, these visualizations reveal that the spaces do not have distinct bounds. This limitation stems from the fact that $C_2(\rho)$ depends not only on the eigenvalues $\{\lambda_k\}$ of the state but also on the eigenvectors through overlaps $|\langle j|\psi_k\rangle|^2$, where $\rho = \sum_k \lambda_k |\psi_k\rangle\langle\psi_k|$. For this reason, we complement this analysis by returning to the entropy pair $(S_2(\rho), S(\rho))$, where both quantities are spectrum-dependent. This allows us to recover both the upper and lower boundaries of entropy-coherence space and facilitates a complete geometric characterization in terms of eigenvalue distributions alone.

*Table 1. Classification of coherence distillation using entropy curves with eigenvalue constraints, normalization, and maximum rank. Table shows the five families of curves in a d-dimensional $S_2(\rho) - S(\rho)$, each defined by distinct eigenvalue ordering and normalization conditions. These curves serve as boundaries or special arcs in coherence analyses of quantum states. $\lambda_e$ represents the equivalent eigenvalue. Note that middle-intermediate curves become upper degenerate curves for the sub-Hilbert space.*

| **Entropy Curve** | **Order constraint** | **Normalization Equation** | **maximum Rank** |
|---|---|---|---|
| Lower-curve qubit states | $\lambda_1 \geq \lambda_2 \geq \{\lambda_i\}_{i=3}^{d} = 0$ | $\lambda_1 + \lambda_2 = 1$ | 2 |
| Lower-Intermediate $m^{th}$ curve with $m = 1,2, \dots, d-3$ | $\{\lambda_i\}_{i=1}^{m+1} = \lambda_e \geq \lambda_{m+2} \geq \{\lambda_i\}_{i=m+3}^{d} = 0$ | $(m+1)\lambda_e + \lambda_{m+2} = 1$ | $m+2$ |
| Lower-Upper curve | $\{\lambda_i\}_{i=1}^{d-1} = \lambda_e \geq \lambda_d \geq 0$ | $(d-1)\lambda_e + \lambda_d = 1$ | $d$ |
| Middle-Intermediate $n^{th}$ curve with $n = 1,2, \dots, d-3$ | $\lambda_1 \geq \{\lambda_i\}_{i=2}^{n+2} = \lambda_e \geq \{\lambda_i\}_{i=n+3}^{d} = 0$ | $\lambda_1 + (n+1)\lambda_e = 1$ | $n+2$ |
| Upper degenerate curve | $\lambda_1 \geq \{\lambda_i\}_{i=2}^{d} = \lambda_e \geq 0$ | $\lambda_1 + (d-1)\lambda_e = 1$ | $d$ |

The dynamics of a quantum system transitioning from a randomly mixed state as shown by a squared marker in Fig. 3 to state with increasing coherence represented by the evolution of the density matrix. We gradually introduce coherence by adding small random off-diagonal elements, with their strength increasing iteratively. By calculating $\mathrm{C}_{l_1}(\rho)$ directly from the generated density matrices and accessing Tsallis-2 $\mathrm{C}_{q=2}(\rho)$ and $\mathrm{C}_r(\rho)$ (and $C_r{}^{\text{cross}}(\rho)$) from $S_2(\rho)$ and $S(\rho)$, respectively, we trace the evolution of quantum states as they transition from a classical (diagonal) state to states with increasing coherence. By utilizing $\mathrm{C}_{l_1}(\rho)$, we can track the evolution of coherence in quantum systems and probe specific regions of the $S_2(\rho) - S(\rho)$ plane without directly calculating the entropies at every step, which can be prohibitively expensive for large quantum systems due to the need to process the entire eigenvalue spectrum. This allows for an efficient characterization of coherence within the plane, where its regions can be explored based on $C_{l_1}(\rho)$, which reflects coherence in a more computationally tractable manner. Thus, we establish a robust framework for characterizing coherence across diverse state configurations, which may provide deeper insights into the role of coherence in quantum state evolution. This approach provides a detailed exploration of how coherence affects key information-theoretic measures, which shed light on the gradual restoration of coherence in quantum systems and its influence on their purity and entropy properties.

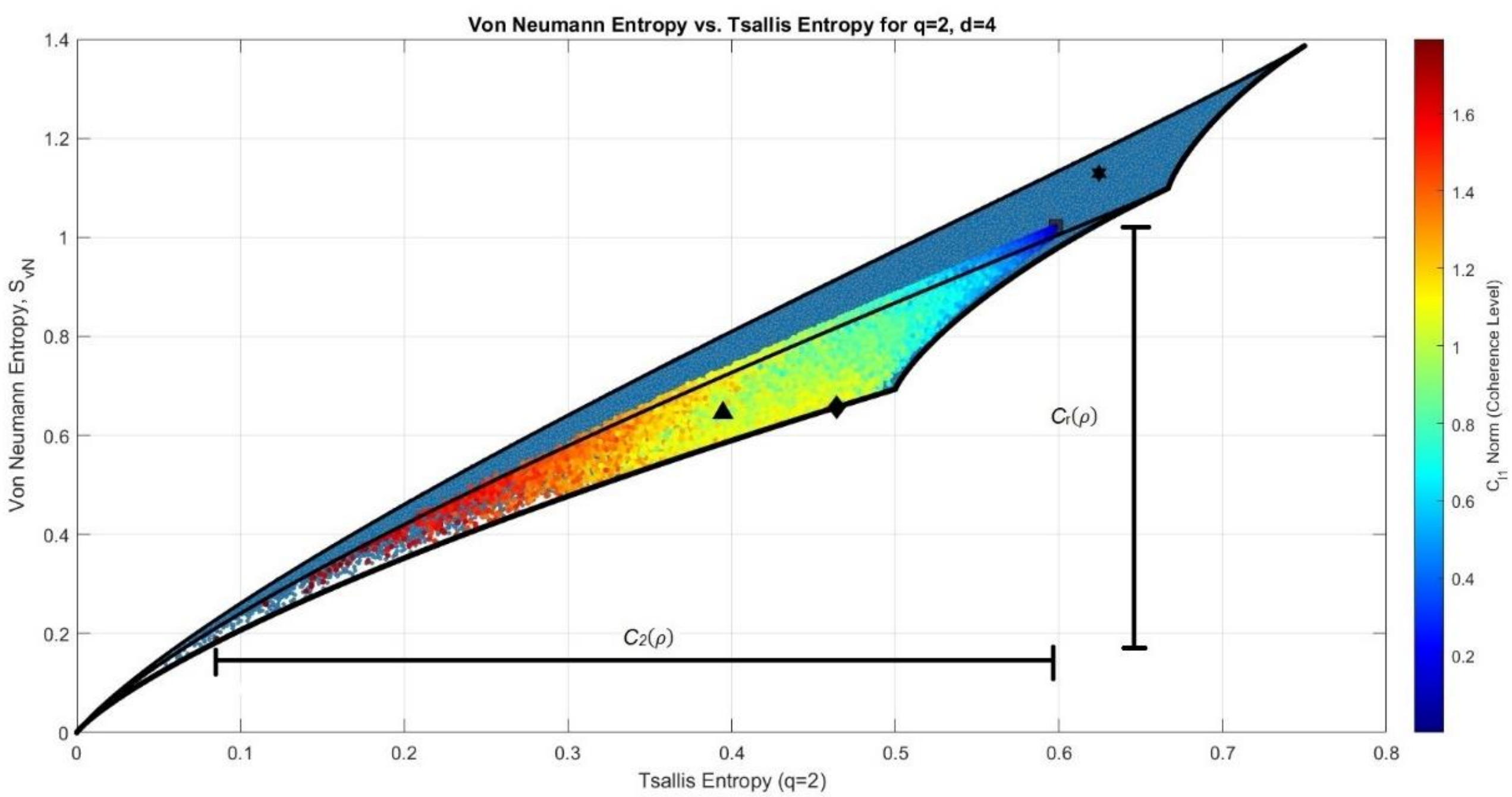

*FIG. 3 illustrates $S_2(\rho) - S(\rho)$ plane for $d = 4$ density matrices of varying ranks. The horizontal and vertical lines outside the region indicates the maximum $C_r(\rho)$ and $C_{q=2}(\rho)$ for a randomly generated matrix shown with a square marker whose transitions across quantum states is shown with color map representing coherence $C_{l_1}(\rho)$.The full-rank matrix (rank 4) is marked with black hexagons, rank 3 with black triangles, and rank 2 with black diamonds, indicating sub-Hilbert spaces. Boundary curves are specified for the coherence distillation such as first-lower and upper curves belong to qubit and degenerate distillations, respectively.*

On the one hand, In Fig. 3, the lower curve delineates the boundary for qubit states on which points belong to the qubit coherence distillation. States on this curve are characterized by eigenvalue distributions $\lambda_1 \geq \lambda_2 \geq \lambda_3 = \lambda_4 = 0$, hence form a rank-2 density operator. Detailed eigenvalue distributions for any $d$-dimensional Hilbert space, which characterize different coherence distillation curves are given in Table 1 and shown in Fig. 3 for $d = 4$. Th distillable coherence in the asymptotic limit,

$$C_d(\rho) = S\left(\rho^{\text{diag}}\right) - S(\rho), \tag{10}$$

quantifies the maximal rate at which copies of $\rho$ can be converted to the qubit state via IO as $n \to \infty$ [6, 50]. The curve indicates that concentrating all the coherence into two nonzero eigenvalues reduces the $S(\rho)$ for a given $S_2(\rho)$, thus marking the minimal-coherence boundary wherein a significant fraction of $\lambda_1$ and $\lambda_2$ contributes directly to qubit-like coherence.

On the other hand, the upper curve in Fig. 3 illustrates degenerate coherence distillation, governed by the constraint $\lambda_1 \geq \lambda_2 = \lambda_3 = \lambda_4$ for the eigenvalues of the density matrix $\rho$. This symmetry maximizes the $S(\rho)$ for a fixed Tsallis entropy $S_2(\rho)$, which traces out the extremal boundary of the entropy-coherence trade-off and provides an operational meaningful measure of purity [50]. We define the degenerate distillable coherence $C_d^{\text{deg}}(\rho)$ as the supremal rate $R$ at which a coherent, non-diagonal target state $\rho_{\text{deg}}$ (with eigenvalues $\lambda_1 \geq \lambda_2 = \cdots = \lambda_d$) can be asymptotically obtained from $\rho^{\otimes n}$ via quantum operation. Concretely, $R$ is achievable if there exists a sequence of free operations $\{\Delta_i\}$ such that, for sufficiently large $n$, the transformed state $\Delta_i[\rho^{\otimes n}]$ approximates ${\rho_{deg}}^{\otimes \lfloor nR \rfloor}$ with arbitrarily small error in the trace norm. Formally:

$$\lim_{n\to\infty} \operatorname*{Inf}_{\Delta_i} \left\| \Delta_i[\rho^{\otimes n}] - ({\rho_{deg}}^{\otimes \lfloor nR \rfloor}) \right\|_1 = 0, \tag{11}$$

where $\lfloor nR \rfloor$ denotes the integer part of $nR$. The degenerate distillable coherence $C_d^{\text{deg}}(\rho)$ is then given by the supremum of such $R$ over all possible free operations $\{\Delta_i\}$, such that

$$C_d^{\text{deg}}(\rho) = \sup\left\{ R \,\middle|\, \lim_{n\to\infty} \operatorname*{Inf}_{\Delta_i} \left\| \Delta_i[\rho^{\otimes n}] - ({\rho_{deg}}^{\otimes \lfloor nR \rfloor}) \right\|_1 = 0 \right\} \tag{12}$$

Here, $\rho^{\otimes n}$ represents $n$ copies of the original state $\rho$, while ${\rho_{deg}}^{\otimes \lfloor nR \rfloor}$ is the tensor product of $\lfloor nR \rfloor$ copies of the target degenerate resource state. The condition that the trace-norm distance goes to zero ensures that, in the limit of infinitely many copies, $\rho^{\otimes n}$ can be faithfully converted into ${\rho_{\text{deg}}}^{\otimes \lfloor nR \rfloor}$. This mirrors the standard structure of resource theories (e.g., entanglement or coherence distillation), where a similar limit-based definition governs the optimal asymptotic rate of generating a target resource. However, the key novelty here lies in choosing $\rho_{\text{deg}}$ with degenerate eigenvalues, rather than a single-qubit maximally coherent state. This choice enables higher-dimensional or symmetrically weighted superpositions to be distilled. This formulation favors states with uniform or nearly uniform eigenvalue distributions, which may offer potential advantages in high-dimensional quantum information tasks by reducing overhead associated with the conversion of multiple state copies into symmetrically weighted coherent superpositions.

However, a careful analysis reveals important nuances when comparing standard distillable coherence $C_d(\rho)$ with degenerate distillable coherence $C_d^{\mathrm{deg}}(\rho)$. Suppose, we distill $k = nC_d^{\mathrm{deg}}(\rho)$ copies of the degenerate state $\rho_{\mathrm{deg}}$ from $n$ copies of the initial state $\rho$. Since coherence cannot increase under free operations, it follows that

$$C_d\left({\rho_{\mathrm{deg}}}^{\otimes k}\right) \leq C_d(\rho^{\otimes n}). \tag{13}$$

Using the additivity property of the relative entropy of coherence, the Eq. (15) becomes

$$kC_d\left(\rho_{\mathrm{deg}}\right) \leq nC_d(\rho). \tag{14}$$

Therefore, coherence monotonicity under free operations, combined with the additivity of the relative entropy of coherence, imposes the bound

$${C_d}^{\mathrm{deg}}(\rho) \leq \frac{C_d(\rho)}{C_d\left(\rho_{\mathrm{deg}}\right)}, \tag{15}$$

where $C_d\left(\rho_{\mathrm{deg}}\right) = S\left(\mathrm{diag}\left(\rho_{\mathrm{deg}}\right)\right) - S(\rho_{\mathrm{deg}})$ is the relative entropy of coherence for the degenerate state, and $\mathrm{diag}\left(\rho_{\mathrm{deg}}\right)$ denotes the diagonal part of $\rho_{\mathrm{deg}}$ in the computational basis. We gratefully acknowledge the reviewer, for emphasizing the importance of this inequality. To illustrate the role of this product inequality ${C_d}^{\mathrm{deg}}(\rho)C_d\left(\rho_{\mathrm{deg}}\right) \leq C_d(\rho)$, consider pure states on the degenerate boundary for $d = 4$, numerical calculations show that $C_d\left(\rho_{\mathrm{deg}}\right) \leq 1$ for eigenvalue distributions with $\lambda_1 \gtrsim 0.812$, which indicates that standard qubit coherence distillation can outperform degenerate coherence distillation in this regime. Conversely, when $\lambda_1 \lesssim 0.812$, one finds $C_d\left(\rho_{\mathrm{deg}}\right) \geq 1$. This highlights an intrinsic trade-off: although degenerate coherence distillation exploits symmetry to achieve potentially more efficient transformations into uniform eigenvalue structures, it may not surpass the fundamental coherence limit set by $C_d(\rho)$.

Nevertheless, the symmetry inherent in degenerate coherence distillation may offer notable advantages for efficient resource transformations in high-dimensional quantum information tasks. Specifically, to target states $\rho_{\mathrm{deg}}$ with uniform or nearly uniform eigenvalue distributions can reduce overhead when we generate symmetrically weighted superpositions. Such states may particularly be advantageous in quantum simulations, metrology, or quantum error correction schemes. Although standard coherence distillation may sometimes outperform degenerate coherence distillation, the added flexibility from degenerate targets facilitates optimal trade-offs between resource consumption and dimensional structure, which may indicate its complementary role within coherence resource theory.

Finally, we extend beyond the usual depiction of eigenvalue distributions in the $S_2(\rho) - S(\rho)$ plane and move into the $S(\rho) - (1 - \lambda_{\max})$ plane. As detailed below, this new perspective reveals a refined, state-dependent entropic uncertainty relation that sometimes surpasses standard bounds. Of course, this parameter space can also yield broader insights into the resource-theoretic properties of quantum coherence. Nevertheless, we now focus on deriving and analyzing a novel entropic uncertainty relation, one that explicitly incorporates the largest eigenvalue of a state that may tighten uncertainty constraints beyond existing limits, particularly in non-Mutually unbiased bases (non-MUBs).

## IV. Refined Entropic Uncertainty via Maximum Eigenvalues

Consider an arbitrary $d$-dimensional quantum state ρ with eigenvalues $\{\lambda_i\}_{i=1}^{d}$ arranged in descending order such that $\lambda_{\max} = \lambda_1 \geq \lambda_2 \geq \cdots \geq \lambda_d$. Numerical evidence (as illustrated in Fig. 4) indicates that the von Neumann entropy $S(\rho)$ consistently exceeds the quantity $1 - \lambda_{\max}$. To rigorously establish this observation for any eigenvalue distribution, we proceed as follows.

We introduce the notation $y \equiv 1 - a$, where $a = \lambda_{\max}(\rho)$ is the largest eigenvalue of the $d$-dimensional quantum state. To understand the boundary conditions clearly, we first consider the scenario of maximum von Neumann entropy. This maximum entropy configuration occurs when the eigenvalues, excluding the largest eigenvalue $a$, are evenly distributed as $\frac{1-a}{d-1}$. Introducing the notation $x \equiv S(\rho)$, we treat $a$ as a parameter spanning $\left[\frac{1}{d}, 1\right]$. This gives

$$x(a) = -\left[a \log a + (1-a) \log\left(\frac{1-a}{d-1}\right)\right], \tag{16}$$

and

$$y(a) = 1 - a. \tag{17}$$

A direct derivative calculation shows that $x(a)$ is strictly decreasing in $a$, whereas $y(a)$ is also decreasing in $a$. Eliminating $a$ thus renders $y$ as an increasing function $y(x)$. Further, since the second derivative $x^{//}(a) < 0$, the map $a \mapsto x(a)$ is strictly concave in $a$, which implies $x \mapsto y(x)$ is strictly convex. Hence $\{(x(a), y(a))\}$ sweeps out a smooth, monotonically increasing boundary curve in the $(S_{vN}, 1-\lambda_{max})$ plane (see Fig. (4)). This curve runs from a (0,0) at $a = 1$ to $(\log d, 1-1/d)$ at $a = 1/d$.

For a fixed maximum eigenvalue $a$, the remaining eigenvalues must satisfy $\sum_{i=2}^{d} \lambda_i = 1 - a$, with each $\lambda_i \leq a$. To minimize $S(\rho)$ under this constraint, the entropy contribution from the eigenvalues $\sum_{i=2}^{d} \lambda_i$ must be minimized by concentrating the remaining probability mass into as few eigenvalues as possible. Specifically, one achieves minimal entropy by assigning the maximal permissible eigenvalue $a$ to as many of the remaining eigenvalues as feasible. Formally, this involves setting $k$ eigenvalues equal to $a$, where $k$ is the largest integer satisfying $k \leq (1-a)/a$, and if necessary, assigning the residual probability mass $b = 1 - a - ka$ to exactly one additional eigenvalue, with all others set to zero.

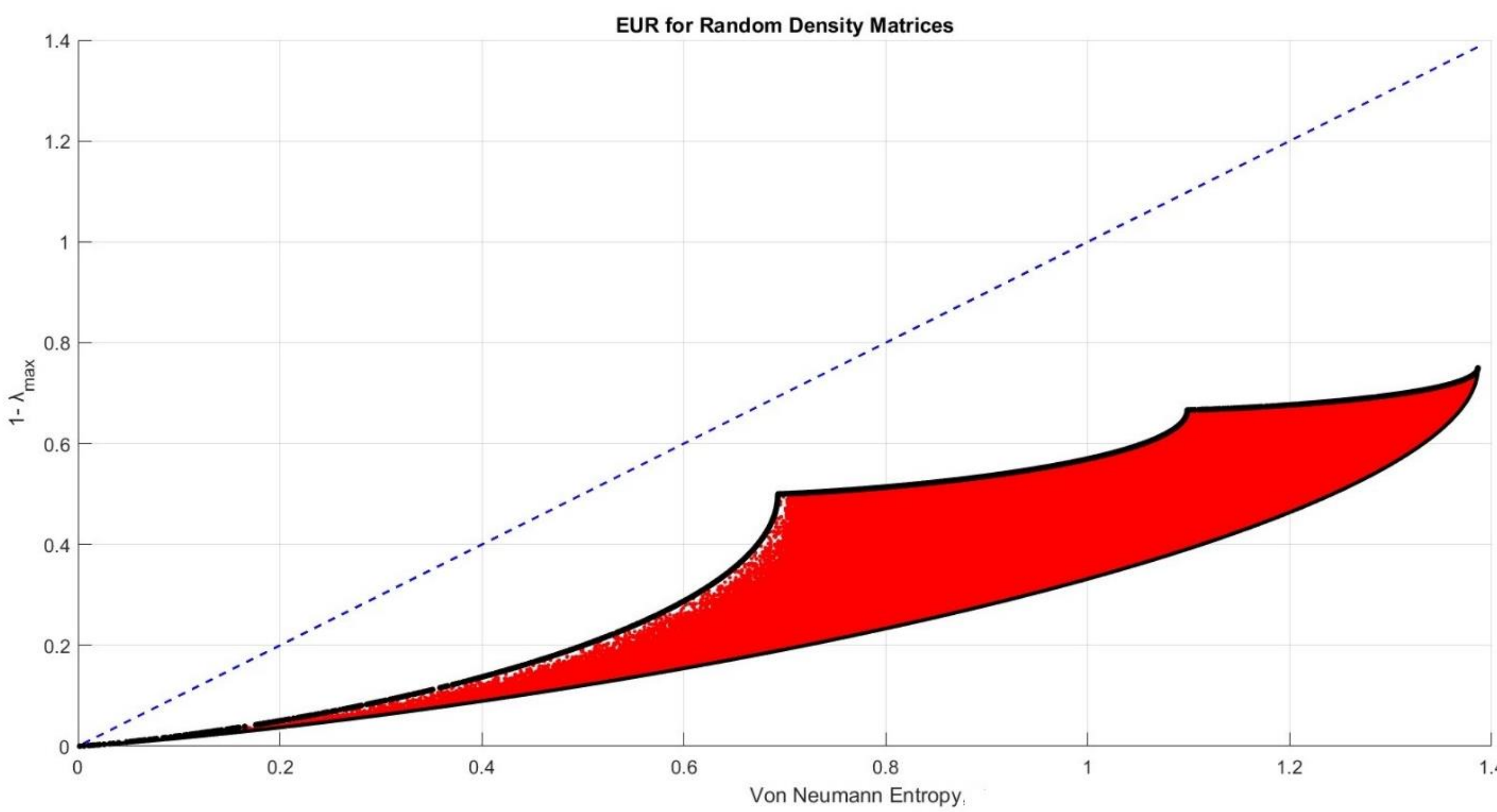

*FIG. 4 explores the relationship between von Neumann entropy $S(\rho)$ and $\sum_{i=2}^{d} \lambda_i = 1 - \lambda_{max}$ for randomly generated density matrices. Data points represent matrices with varying eigenvalue distributions, and a reference line $y = x$ indicates equality which is always above the bound for any dimension and shown here for $d = 4$.*

With this optimized eigenvalue distribution, the minimal achievable entropy $S_{\min}(a)$ is thus

$$S_{\min}(a) = -a \log a - ka \log a - b \log b, \text{ if } b > 0, \quad (18)$$

and reduces to $S_{\min}(a) = -(k+1)a \log a = -\log a$ if the residual probability $b = 0$. Importantly, it can be shown that for all $a \in [1/d, 1]$, the inequality $-\log a \geq 1 - a$ holds true.

This follows from analyzing the function $g(a) = -\log a - (1-a)$, for which one finds $g(1) = 0$ and for $a < 1$, $g^{/}(a) = -\frac{1}{a} + 1 \leq 0$, which indicates that $g(a) \geq 0, \forall\, a \in [0,1]$. Since $S(\rho) \geq S_{\min}(a) \geq -\log a \geq 1 - a$, we establish the general entropic bound,

$$S(\rho) \geq 1 - \lambda_{\max}(\rho), \quad \forall\, \rho. \tag{19}$$

This result, valid for any eigenvalue distribution, thus rigorously confirms the numerical trends depicted in Fig. 4 where the reference blue line shows that any point located on the plane satisfies, $S(\rho) \geq 1 - \lambda_{\max}$, thereby reveals a fundamental, state-dependent constraint between entropy and $\lambda_{\max}$.

Furthermore, we extend the framework to measurement outcomes. Define $\lambda^{j}{}_{\max}(\rho)$, $j = X, Z$ as the largest eigenvalue of the post-measurement state in basis $j$ (for instance, $j = X$ or $j = Z$). Substituting the corresponding Shannon entropies $H(j,\rho)$, for $S(\rho)$, one obtains

$$H(X,\rho) + H(Z,\rho) \geq 2 - [\lambda^{X}{}_{\max}(\rho) + \lambda^{Z}{}_{\max}(\rho)]. \tag{20}$$

Indeed, this can be extended to $N$ number of measurement bases. In contrast to standard Maassen-Uffink (MU) uncertainty bound [45], which depends on only the overlap of the two bases, this state-dependent relation tightens whenever $\lambda^{j}{}_{\max}(\rho)$ is sufficiently small. Therefore, Eq. (20) is more refined quantum-information constraint. Thus, replacing the overlap term $-2\log c$ in MU by a combination of $1 - \lambda_{\max}(X) + 1 - \lambda_{\max}(Z)$ yields an operational entropic bound that is sensitive to the actual state $\rho$ under consideration, rather than merely the geometry of the two measurement bases.

Following the general framework established in Eq. (20), we now illustrate its application with two examples: a qubit system ($d = 2$) and a qudit system ($d = 4$). For the qubit case, consider a mixed state defined as $\rho = p|\psi\rangle\langle\psi| + (1-p)\, I/2$, $I$ being the identity matrix. We assume $p = 0.8$ and take $\phi = \pi/8$ for initial state $|\psi\rangle = \cos\phi|0\rangle + \sin\phi\,|1\rangle$. This yields a density matrix $\rho \approx \begin{pmatrix} 0.7828 & 0.2828 \\ 0.2828 & 0.2172 \end{pmatrix}$. Measurements are performed in the computational basis $X = \{|0\rangle, |1\rangle\}$ and rotated basis $Z = \{|z_0\rangle, |z_1\rangle\}$, with $|z_0\rangle = \cos\phi\,|0\rangle + \sin\phi\,|1\rangle$ and $|z_1\rangle = -\sin\phi\, 0\rangle + \cos\,\phi\,|1\rangle$. The maximum overlap $c \approx 0.9239$ leading to a MU bound of $H(X,\rho) + H(Z,\rho) \geq -2\ln c \approx 0.1584$ nats. The largest eigenvalue-based EUR uses $\lambda^{X}{}_{max}(\rho) \approx 0.7828$ and $\lambda^{X}{}_{max}(\rho) = 0.90$, which gives a tighter bound of 0.3172 nats. The actual entropy sum $H(X,\rho) + H(Z,\rho) \approx 0.848$ nats, which surpasses both bounds but aligns more closely with the eigenvalue-based EUR, which highlights it advantage in capturing state-specific uncertainty for non-mutually unbiased bases (non-MUBs).

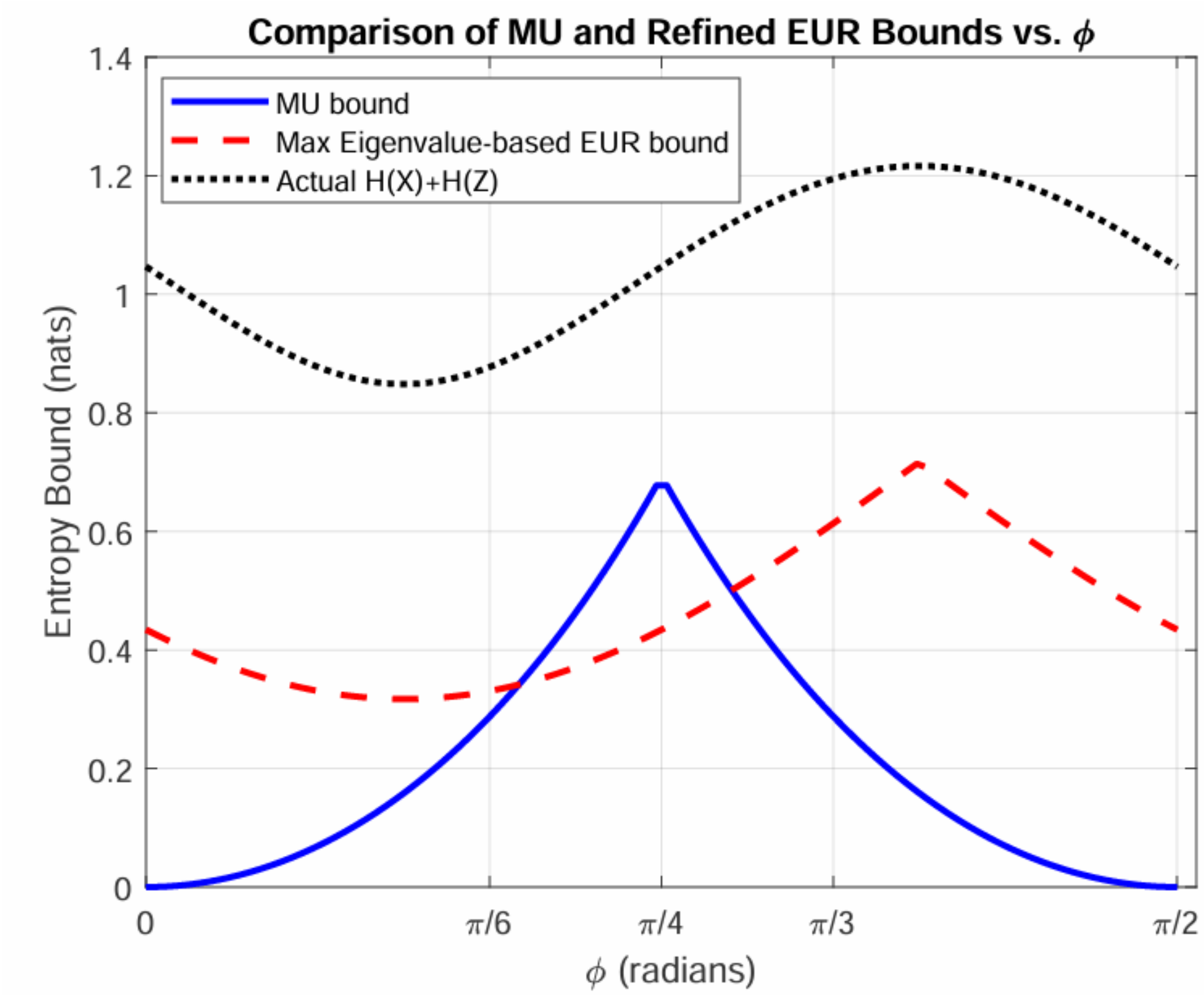

*FIG. 5 Comparison of the MU bound, largest eigenvalue-based entropic uncertainty bound, and the actual entropy sum H(X)+H(Z) as a function of rotation angle $\phi$ for a qubit mixed state. The refined bound consistently outperforms the MU bound, except near mutually unbiased configurations.*

For the qudit case, we examine a state $\rho = p|\psi\rangle\langle\psi| + (1-p)\, I/4$, with $p = 0.75$ and $|\psi\rangle = \frac{1}{2}(|0\rangle + |1\rangle + |2 + |2\rangle)$. The density matrix is $\rho = 0.8175J + 0.0625I$, with $J$ the $4 \times 4$ all-ones matrix, resulting in diagonal elements $\rho_{ii} = 0.25$ and off-diagonal elements $\rho_{ij} = 0.1875$ for $i \neq j$. Measurements occur in the computational basis $X = \{|0\rangle, |1\rangle, |2\rangle, |3\rangle\}$ and a rotated basis $Z = \{|z_0\rangle, |z_1\rangle, |z_2\rangle, |z_3\rangle\}$, where $|z_0\rangle = 0.8|0\rangle + 0.6|1\rangle$, $|z_1\rangle = -0.6|0\rangle + 0.8|1\rangle$, $|z_2\rangle = 0.8|2\rangle + 0.6|3\rangle$, and $|z_3\rangle = -0.6|2\rangle + 0.8|3\rangle$. With the maximum overlap of $c = 0.8$, the MU bound is $-2\ln c \approx 0.446$ nats. The largest eigenvalue-based EUR, with $\lambda^X{}_{max}(\rho) = 0.25$ and $\lambda^X{}_{max}(\rho) = 0.43$, provides $2 - (0.25 + 0.43) = 1.32$, a notably stricter constraint. The actual entropy sum is approximately 2.483 nats, again closer to the new bound. This demonstrates the

efficiency of the largest eigenvalue-based bound in higher dimensional systems where spectral properties significantly enhance uncertainty estimates.

To explore these bounds, a simulation of the qubit case is shown in Fig. (5) by varying the $Z-$basis rotation angle $\phi$ from 0 to $\pi/2$. For the state $\rho$ defined in the example 1 (above), the $X-$ basis remains fixed as $\{|0\rangle, |1\rangle\}$, while the $Z-$ basis is $\{|z_0\rangle = [\cos\phi; \sin\phi], |z_1\rangle = [-\sin\phi; \cos\phi]\}$. The overlap $c = \max\{\langle x_i | z_j \rangle\}$, the MU bound $-2\ln c$, the eigenvalue-based bound $2 - (\lambda^X{}_{\max} + \lambda^Z{}_{\max})$, and the actual entropy sum $H(X) + H(Z)$ across $\phi$ values is computed. Plotting these quantities reveals that the MU bound varies with $c$ (and thus $\phi$), while the eigenvalue-based bound adjusts to the state's maximum probabilities, which consistently offers a tighter constraint, especially when bases deviate from mutual unbiasedness. Fig. 5 clearly demonstrates that the maximum eigenvalue-based EUR offers a tighter, more state-sensitive bound than the MU bound over a range of rotation angles $\phi$. The only region where the MU bound approaches or temporarily exceeds the eigenvalue-based bound is near the angle $\phi \approx \pi/4$, which corresponds to a MUBs for qubit systems (i.e., when the $Z-$basis is rotated by 45° from the $X-$basis). In this scenario, the overlap $c$ approaches $1/\sqrt{2}$, which is the condition for MUBs in dimension 2.

This visualization underscores the eigenvalue-based EUR's sensitivity to spectral properties, which provides a practical tool for analyzing uncertainty across measurement settings in quantum information tasks.

To sum up, the eigenvalue-based EUR enhances the traditional MU bound by integrating state-dependent maximum probabilities $\lambda^j{}_{\max}$, as evidenced in both qubit and qudit examples and supported by numerical simulation. Its tighter bounds are particularly valuable in higher dimensions and non-mutually unbiased scenarios, closely tracking actual entropy sums. For future work, a hybrid bound such as $\max(-\ln c\,, 2 - (\lambda^X{}_{\max} + \lambda^Z{}_{\max}))$ could be explored, combining the eigenvalue-based EUR's state-specific precision with the MU bound's robustness, potentially offering a versatile uncertainty metric adaptable to diverse quantum systems and experimental conditions.

## V. Conclusions

In this work we present a geometric approach to quantum coherence by constructing entropic diagrams based on von Neumann and Tsallis-2 entropies. We derived relative entropy of coherence by using the Schur–Horn majorization theorem and introduced a new coherence monotone, relative cross-entropy of coherence. We analyzed its behavior under a class of quantum operations, including DIO and SIO, and established that it satisfies monotonicity under these operations. Moreover, we proved that the measure satisfies convexity and additivity under IO. While the measure diverges for pure coherent states and is therefore restricted to mixed-state scenarios, it offers a meaningful perspective on coherence loss and may serve as a useful tool for characterizing coherence degradation in quantum information tasks.

We then employed entropic diagram to characterize quantum coherence. These diagrams can reveal distinct entropy–coherence distillation curves that are generated from different eigenvalue configurations, which we interpret as distillation pathways. To address scenarios which may require high-dimensional or symmetrically structured coherence, we introduced the notion of degenerate coherence distillation. Unlike conventional distillation that targets qubit-like maximally coherent states, our approach considers structured target states with partially uniform spectra. These lie along the upper boundary of the entropy diagram and enable alternative strategies for coherence concentration. More specifically, it may be useful in qudit protocols or multi-mode optical systems. While degenerate distillation may not always outperform qubit-based schemes in asymptotic rate, it can provide a complementary pathway that aligns with emerging demands in high-dimensional quantum technologies.

From the maximum eigenvalue−entropic diagram, we derived a refined entropic uncertainty relation that depends explicitly on the maximum eigenvalue of the state. This new bound improves upon the conventional Maassen–Uffink inequality by providing state-dependent constraints that more accurately capture the interplay between entropy and measurement uncertainty for non-mutually unbiased (non-MUBs). Numerical and analytical investigations show that this relation yields tighter bounds in both low− and high-dimensional settings, and is particularly effective in non-MUBs measurement scenarios.

Altogether, this work unifies spectral analysis, resource quantification, and geometric visualization, which may provide new tools for understanding and applying quantum coherence. The results established here using entropic diagrams may not only as an insightful visualization method but they can also provide a rigorous foundation to derive entropic uncertainty bounds and to guide coherence-based distillation protocols. Future directions include extending this framework to multi-basis entropic uncertainty relations, which may explore analogous constructions for other quantum resources such as magic or non-Markovianity, and to apply these insights to quantum technologies where structured coherence plays a central role. Together, these results demonstrate how entropic diagrams and majorization structure can be used not only to quantify coherence but also to visualize resource trade-offs, design alternative distillation schemes, and derive improved bounds on quantum uncertainty. We anticipate that this work may find broader applications in quantum metrology, communication, and coherence-based quantum technologies.

## ACKNOWLEDGMENTS

This work was supported by the National Natural Science Foundation of China (Grant Nos. 12475009, 12075001, and 61601002), Anhui Provincial Key Research and Development Plan (Grant No. 2022b13020004), Anhui Province Science and Technology Innovation Project (Grant No. 202423r06050004), and Anhui Provincial University Scientific Research Major Project (Grant No. 2024AH040008).

## References:

1. R. J. Glauber, Coherent and incoherent states of the radiation field, Phys. Rev. **131**, 2766 (1963).
2. E. C. G. Sudarshan, Equivalence of semiclassical and quantum mechanical descriptions of statistical light beams, Phys. Rev. Lett. **10**, 277 (1963).
3. E. Wolf, Introduction to the Theory of Coherence and Polarization of Light (Cambridge University Press, Cambridge, 2007).
4. J. Åberg, arXiv:quant-ph/0612146 (2006).
5. T. Baumgratz, M. Cramer, and M. B. Plenio, Quantifying coherence, Phys. Rev. Lett. **113**, 140401 (2014).
6. A. Winter and D. Yang, Operational resource theory of coherence, Phys. Rev. Lett. **116**, 120404 (2016).
7. B. Yadin, J. Ma, D. Girolami, M. Gu, and V. Vedral, Quantum processes which do not use coherence, Phys. Rev. X **6**, 041028 (2016).
8. E. Chitambar and G. Gour, Critical examination of incoherent operations and a physically consistent resource theory of quantum coherence, Phys. Rev. Lett. **117**, 030401 (2016).

9. E. Chitambar, Dephasing-covariant operations enable asymptotic reversibility of quantum resources, Phys. Rev. A **97**, 050301(R) (2018).
10. I. Marvian and R. W. Spekkens, How to quantify coherence: Distinguishing speakable and unspeakable notions, Phys. Rev. A **94**, 052324 (2016).
11. I. Marvian, R. W. Spekkens, and P. Zanardi, Quantum speed limits, coherence, and asymmetry, Phys. Rev. A **93**, 052331 (2016).
12. J. I. de Vicente and A. Streltsov, Genuine quantum coherence, J. Phys. A: Math. Theor. **50**, 045301 (2017).
13. E. Chitambar and M. H. Hsieh, Relating the resource theories of entanglement and quantum coherence, Phys. Rev. Lett. **117**, 020402 (2016).
14. M. Hillery, Coherence as a resource in decision problems: The Deutsch-Jozsa algorithm and a variation, Phys. Rev. A **93**, 012111 (2016).
15. G. Gour, Role of quantum coherence in thermodynamics, PRX Quantum **3**, 040323 (2022).
16. J. Goold, M. Huber, A. Riera, L. del Rio, and P. Skrzypczyk, The role of quantum information in thermodynamics—a topical review, J. Phys. A: Math. Theor. **49**, 143001 (2016).
17. D. P. Pires, I. A. Silva, E. R. de Azevedo, D. O. Soares-Pinto, and J. G. Filgueiras, Coherence orders, decoherence, and quantum metrology, Phys. Rev. A **98**, 032101 (2018).
18. I. Marvian and R. W. Spekkens, Modes of asymmetry: The application of harmonic analysis to symmetric quantum dynamics and quantum reference frames, Phys. Rev. A **90**, 062110 (2014).
19. C. Zhang, T. R. Bromley, Y. F. Huang, H. Cao, W. M. Lv, B. H. Liu, *et al.*, Demonstrating quantum coherence and metrology that is resilient to transversal noise, Phys. Rev. Lett. **123**, 180504 (2019).
20. M.-L. Hu, X. Hu, J. Wang, Y. Peng, Y.-R. Zhang, and H. Fan, Quantum coherence and geometric quantum discord, Phys. Rep. **762**, 1 (2018).
21. A. Streltsov, G. Adesso, and M. B. Plenio, Colloquium: Quantum coherence as a resource, Rev. Mod. Phys. **89**, 041003 (2017).
22. H. Yamasaki, M. K. Vijayan, and M. H. Hsieh, Hierarchy of quantum operations in manipulating coherence and entanglement, Quantum **5**, 480 (2021).

23. S. Du, Z. Bai, and Y. Guo, Conditions for coherence transformations under incoherent operations, Phys. Rev. A **91**, 052120 (2015).
24. E. Chitambar and G. Gour, Comparison of incoherent operations and measures of coherence, Phys. Rev. A **94**, 052336 (2016).
25. E. Chitambar and G. Gour, Erratum: Comparison of incoherent operations and measures of coherence [Phys. Rev. A **94**, 052336 (2016)], Phys. Rev. A **95**, 019902(E) (2017).
26. H. Zhu, Z. Ma, Z. Cao, S. M. Fei, and V. Vedral, Operational one-to-one mapping between coherence and entanglement measures, Phys. Rev. A **96**, 032316 (2017).
27. Y. Alvarez, M. Losada, M. Portesi, and G. M. Bosyk, Complementarity between quantum coherence and mixedness: A majorization approach, Commun. Theor. Phys. **75**, 055102 (2023).
28. G. M. Bosyk, M. Losada, C. Massri, H. Freytes, and G. Sergioli, Generalized coherence vector applied to coherence transformations and quantifiers, Phys. Rev. A **103**, 012403 (2021).
29. K. Fang, Distillation and simulation in quantum information (Doctoral dissertation, University of Technology Sydney, 2018).
30. L. Lami, B. Regula, and G. Adesso, Generic bound coherence under strictly incoherent operations, Phys. Rev. Lett. **122**, 150402 (2019).
31. B. Regula, K. Fang, X. Wang, and G. Adesso, One-shot coherence distillation, Phys. Rev. Lett. **121**, 010401 (2018).
32. K. Fang, X. Wang, L. Lami, B. Regula, and G. Adesso, Probabilistic distillation of quantum coherence, Phys. Rev. Lett. **121**, 070404 (2018).
33. C. L. Liu and D. L. Zhou, Deterministic coherence distillation, Phys. Rev. Lett. **123**, 070402 (2019).
34. E. Chitambar, A. Streltsov, S. Rana, M. N. Bera, G. Adesso, and M. Lewenstein, Assisted distillation of quantum coherence, Phys. Rev. Lett. **116**, 070402 (2016).
35. P. Char, D. Chakraborty, A. Bhar, I. Chattopadhyay, and D. Sarkar, Catalytic transformations in coherence theory, Phys. Rev. A **107**, 012404 (2023).
36. A. Anshu, M. H. Hsieh, and R. Jain, Quantifying resource in catalytic resource theory, Phys. Rev. Lett. **115**, 150502 (2015).

37. N. Shiraishi and R. Takagi, Arbitrary amplification of quantum coherence in asymptotic and catalytic transformation, Phys. Rev. Lett. **132**, 180202 (2024).
38. P. Harremoës and F. Topsøe, Inequalities between entropy and index of coincidence derived from information diagrams, IEEE Trans. Inf. Theory **47**, 2944 (2001).
39. P. Harremoës and F. Topsøe, Information diagrams: Entropy, index of coincidence and probability of error, in Proc. IEEE Int. Symp. Inf. Theory (2001).
40. A. Aiello and J. P. Woerdman, Physical bounds to the entropy-depolarization relation in random light scattering, Phys. Rev. Lett. **94**, 090406 (2005).
41. T. Aziz, P. Li, D. Chen, D. Lv, and H. Ma, Physically realizable space for the purity-depolarization plane for polarized light scattering media, Phys. Rev. Lett. **119**, 033202 (2017).
42. T. Aziz, H. He, P. Li, and H. Ma, Purity-depolarization relations and the components of purity of a Mueller matrix, Opt. Express 27, 22645 (2019).
43. E. Rastegin, Uncertainty relations in terms of generalized entropies derived from information diagrams, arXiv:2305.18005 (2023).
44. S. Huang, H. L. Yin, Z. B. Chen, and S. Wu, Entropic uncertainty relations for multiple measurements assigned with biased weights, Phys. Rev. Res. 6, 013127 (2024).
45. H. Maassen and J. B. Uffink, Generalized entropic uncertainty relations, Phys. Rev. Lett. 60, 1103 (1988).
46. M. Atiyah, Convexity and commuting Hamiltonians, Bull. Lond. Math. Soc. 14, 1 (1982).
47. J. Hilgert, K. H. Neeb, and W. Plank, Symplectic convexity theorems, Sem. Sophus Lie 3, 123 (1993).
48. M. A. Nielsen, Lecture notes, Department of Physics, University of Queensland, Australia (2002).
49. V. Vedral and M. B. Plenio, Entanglement measures and purification procedures, Phys. Rev. A 57, 1619 (1998).
50. E. Rastegin, Quantum-coherence quantifiers based on the Tsallis relative α entropies, Phys. Rev. A **93**, 032136 (2016).
51. H. Zhao and C.-S. Yu, Coherence measure in terms of the Tsallis relative α entropy, Sci. Rep. **8**, 299 (2018).
52. Z. Xi and S. Yuwen, Epsilon-smooth measure of coherence, Phys. Rev. A **99**, 012308 (2019).

53. K. Bu, U. Singh, S. M. Fei, A. K. Pati, and J. Wu, Max-relative entropy of coherence: An operational coherence measure, Phys. Rev. Lett. **119**, 150405 (2017).
54. G. Saxena, E. Chitambar, and G. Gour, Dynamical resource theory of quantum coherence, Phys. Rev. Research **2**, 023298 (2020).